# Multivariate dependence and genetic networks inference

Adam A. Margolin[1,2], Kai Wang[2,3], Andrea Califano[2], Ilya Nemenman[2,4]

[1]Cancer Program, The Broad Institute of Harvard and MIT, Cambridge, MA 02142, USA

[2]Joint Centers for Systems Biology, Columbia University Medical Center, New York, NY 10032, USA

[3]Oncology Research Unit, Pfizer, San Diego, CA 92121, USA

[4]Departments of Physics and Biology, Computational and Life Sciences Strategic Initiative, Emory University, Atlanta, GA 30322, USA

## Abstract

**Motivation:** A critical task in systems biology is the identification of genes that interact to control cellular processes by transcriptional activation of a set of target genes. Many methods have been developed to use statistical correlations in high-throughput datasets to infer such interactions. However, cellular pathways are highly cooperative, often requiring the joint effect of many molecules, and few methods have been proposed to explicitly identify such higher-order interactions, partially due to the fact that the notion of multivariate statistical dependency itself remains imprecisely defined.

**Results:** We define the concept of dependence among multiple variables using maximum entropy techniques and introduce computational tests for their identification. Synthetic network results reveal that this procedure uncovers dependencies even in undersampled regimes, when the joint probability distribution cannot be reliably estimated. Analysis of microarray data from human B cells reveals that third-order statistics, but not second-order ones, uncover relationships between genes that interact in a pathway to cooperatively regulate a common set of targets.

**Contact:** margolin@broadinstitute.org, ilya.nemenman@emory.edu

# 1. Introduction

Reverse engineering molecular interaction networks is a critical challenge in modern systems biology [1]. High-throughput technologies allow simultaneous measurements of the concentrations of thousands of molecular species in a biological system, such as mRNA [2], microRNA [3], proteins [4] and metabolites [5]. Each such experiment may be treated as an observation from a joint probability distribution (JPD), and it is believed that statistical dependencies in this JPD provide clues about biochemical interactions among the species [6]. Thus identifying dependencies in JPDs is an essential task for network reverse engineering, and this problem also is ubiquitous in other branches of systems biology [7-9], as well as in many other applications.

It is clearly understood [10] that statistical dependencies can be characterized by their order (that is, by the number of variables—molecular species—participating in them). Until recently, most network reverse engineering work focused on second-order (pairwise) dependencies. Their identification from data is now a common exercise. In particular, direct (irreducible) interactions can be disambiguated from indirect ones (e.g., two biochemical species correlated due to a common regulator) [11, 12]. However, combinatorial regulation, where multiple effectors combine to regulate a target gene, is prevalent in higher eukaryotes [13]. Correspondingly, recent years have seen a surge in the use of high-throughput data to identify these higher-order structures [6, 14-19]. However, as described below, there has been little work to rigorously define the mathematical basis of the identified multivariate statistical dependencies and the structure of uncovered interactions (e.g., cooperative versus independent regulation). For example, consider two transcription factors, $TF_1$ and $TF_2$, that may regulate the expression of a target gene, T, in different ways, including, but not limited to (note that we use Roman characters to denote gene names and italic ones for gene expressions):

$$\frac{dT}{dt} = h_1(TF_1) + h_2(TF_2) - rT, \qquad (1.1)$$

$$\frac{dT}{dt} = h_1(TF_1)h_2(TF_2) - rT. \qquad (1.2)$$

Here $h_i$'s are single-effector activation terms, such as Hill functions, and $rT$ is the first-order degradation. The first of these equations describes independent activation of the target. In the second equation, both transcription factors act synergistically, for example, due to formation of a transcriptional complex (this type of dependency also applies in the case of a signaling molecule that post-translationally modifies a transcription factor, influencing its ability to regulate the target). We expect $T$ to be statistically dependent on $TF_1$ and $TF_2$ in both cases; however, clearly, there is a difference, since for Eq. (1.2) the effects of $TF_1$ and $TF_2$ on $T$ cannot be studied in isolation from each other, forming a third-order dependency among the variables. With Eqs. (1.1), (1.2) infused with the usual Gaussian noise with variance $\sigma^2$, the resulting steady state equations are

$$P(T \mid TF_1, TF_2) \propto e^{-\frac{1}{2r^2\sigma^2}[T - h_1(TF_1) - h_2(TF_2)]^2}, \qquad (1.3)$$

$$P(T \mid TF_1, TF_2) \propto e^{-\frac{1}{2r^2\sigma^2}[T - h_1(TF_1)h_2(TF_2)]^2}. \qquad (1.4)$$

Thus joint regulation involves a term that couples all three variables in the exponent of the JPD. A reasonable tool for statistical analysis of multivariate interaction patterns should distinguish such high-order structures from additive pairwise interactions, as in Eq. (1.1).

This is a nontrivial task since, even now, there is no consensus definition of an interaction in the multivariate setting. For example, standard statistical methods

[20, 21] introduce many specialized dependence concepts applicable in restricted contexts, such as normal noise, binary, bivariate, or metric data, etc. Alternatively, contingency tables literature associates interactions with deviations of the number of observed counts from their expectations under various independence assumptions [22-24]. Unfortunately, this is limited to categorical data and confounds the definition of dependence with sampling issues. In information theory [25, 26], one can treat continuous and categorical data uniformly [27, 28] and define dependencies based on distributions rather than counts, but none of the information theoretic interaction measures [10, 18, 19, 29-34] have become universally accepted either.

In the context of systems biology, multivariate dependencies have been studied traditionally [6] using probabilistic graphical models [35], such as Bayesian Networks (BNs) or Markov Networks, also known as Markov Random Fields (MRFs). However, these models are generically unable to disambiguate different types of regulation, such as in Eqs. (1.1), (1.2) [36]. This limitation arises from relying on the notion of conditional (in)dependence rather than providing a precise definition of statistical dependency among subsets of variables (see below for more details). That is, many different interaction patterns can give rise to the same conditional independence structure in a MRF.

In this work, we build on the definition of connected interactions proposed by [10] to rigorously define a multivariate statistical interaction. The approach is initially motivated by information theoretic concepts, and it is described in Sec. 2. In Sec. 3 we describe the method in terms of specially adapted factor graph models that generalize BNs and MRFs. We apply the method to a simple synthetic model in Sec. 4 and to a biological dataset from human B cells in Sec. 5. The synthetic model demonstrates the method's ability to infer interactions even for undersampled distributions. For application to biological data, we derive a computationally efficient simplification of the formula for third-order dependencies, and hint at the ability to disambiguate between independent and cooperative regulation.

## 2. Definition of Multivariate Dependence

For two variables, $X_1$ and $X_2$, independence is well defined via decomposition of the bivariate JPD, $P(X_1, X_2) = P(X_1)P(X_2)$, and mutual information $I(X_1; X_2) \equiv I_P(X_1; X_2) \equiv \langle \log_2 P(X_1, X_2) / [P(X_1)P(X_2)] \rangle$ is the unique measure of dependence [26]. Similarly, the total interaction (that is, the deviation from independence) in a multivariate JPD, $P(\{X_i\})$, $i = 1,...,N$, can be measured by the multi-information [33]

$$I_P(X_1;...;X_N) \equiv D_{\text{KL}}[P \| P^*] = \left\langle \log_2 \frac{P(X_1,...,X_N)}{P(X_1)\cdots P(X_N)} \right\rangle_P \tag{1.5}$$

which assigns a specific number of *bits* to the union of all interactions among the studied variables. Here $D_{\text{KL}}$ is the Kullback-Leibler (KL) divergence [37] between the full JPD, $P(X_1,...,X_N)$, and its approximation under the independence assumption, $P^*(X_1,...,X_N) = P(X_1)\cdots P(X_N)$. In order to define multivariate statistical dependence, we seek to partition the total deviation from independence into contributions from interactions among various variable subsets (*specific* pairs, triplets, etc.), and a nonzero contribution from a subset would indicate an interaction among its members.

We first note that $P^*$ is the maximum entropy (MaxEnt) distribution [38, 39] that has the same marginals as $P$ but introduces no statistical dependencies among the variables [10, 32, 40]. Thus the multi-information is the KL divergence between the JPD and its MaxEnt approximation with marginal constraints, and it measures the gain in information by knowing the complete JPD versus assuming total independence. Similarly, MaxEnt distributions consistent with various multivariate marginals of the JPD introduce no statistical interactions beyond those in the said marginals. Thus by comparing the JPD to its MaxEnt approximations under various marginal constraints, one can separate dependencies included in the low-order statistics from those not present in them [32, 40-43].

Specifically, one can define *connected* interactions of a given order, i.e., the interactions that need, at least, the full set of marginals of this order to be captured. Following [10], suppose that we have a network of $N$ variables and we know a set of marginal distributions of all variable subsets of size $k \geq 1$, so that $P(X_{i_1},...,X_{i_k}) = \sum_{X \notin \bigcup_{j=1}^{k} X_{i_j}} P(\{X\})$ is specified. One can ask what is the JPD $P^{(k)}$ that captures all multivariate interactions prescribed by these marginals, but introduces no additional dependencies. That is, one searches for a distribution $P^{(k)}$ with a minimum $I_{P^{(k)}}$ (or, alternatively, with the maximum entropy—MaxEnt—$H_{P^{(k)}}$) such that the constraints $C_{i_1,...,i_k}(P^{(k)}, P) \equiv P^{(k)}(X_{i_1},...,X_{i_k}) - P(X_{i_1},...,X_{i_k}) = 0$ are satisfied.[1] This is given by the MaxEnt, or minimum multi-information, problem [10, 38, 40]:

$$P^{(k)} \equiv \arg\max_{P',\{\lambda\}} \{H[P'] - \sum_{M \in \Omega} \lambda_M C_M(P', P)\}, \qquad (1.6)$$

where $M$'s are sets of constrained variables, such as $M_{i_1,...,i_k} = \{X_{i_1},...,X_{i_k}\}$, and $\Omega = \bigcup M$. Further, $\lambda$'s are the Lagrange multipliers that enforce the marginal

---

[1]All JPDs constrained by the same marginals are said to form a Fréchet class 21.

Joe, H., *Multivariate models and dependence concepts*. 1997, Boca Raton: Chapman and Hall.. For metric variables and simple constraints, these classes are well studied. We know parametric forms for some of them, can check if the constraints are compatible, and if they determine the JPD uniquely.

constraints. They are matrices of the same dimensionality as the constraints they enforce, but we do not write out the indices of JPDs and $\lambda$'s explicitly.

The solution of a MaxEnt problem with marginal constraints has the form of a product of terms dependent on the constrained variables [44]. In particular, for Eq. (1.6),

$$P^{(m)} = \frac{1}{Z} \prod_{i_1 < \cdots < i_m} \psi(M_{i_1 \ldots i_m}), \psi \geq 0, \qquad (1.7)$$

where $Z$ is the normalization and each $\psi$ is a different function, known as a potential, which is determined implicitly by the marginal constraints. In general, no analytical solution for the $\psi$'s exists. However, an algorithm called the *iterative proportional fitting procedure* (IPFP) [45], which iteratively adjusts a trial solution to satisfy each of the constraints in turn, converges to the true solution [44]. The connected information of order $k$ is then

$$I_C^{(k)}(X_1; \ldots; X_N) \equiv D[P^{(k)} \| P^{(k-1)}] = I_{P^{(k-1)}}(X_1; \ldots; X_N) - I_{P^{(k)}}(X_1; \ldots; X_N). \qquad (1.8)$$

This characterizes the increase in information by knowing all marginals of order $k$, as opposed to all marginals of order $k-1$. Note that the multi-information can be decomposed into a series of connected informations, $I_P = \sum_{k=2}^{N} I_C^{(k)}$.

While appealing, the connected interaction construction assigns interaction bits to a particular interaction order. We need to refine the approach to instead assign the bits to a particular combination of variables within this order, which has not yet been done.

To localize (connected) interactions to particular sets of covariates, we note that mutual, multi, and connected information are special cases of a general principle of evaluating the KL divergence between the MaxEnt distributions constrained by a set of marginals and a subset of these marginals (or, alternatively, the difference of entropies of these two MaxEnt distributions, or the negative difference of the multi-informations). If the divergence is positive, then the extra marginal constraints correspond to a nonzero interaction. Thus to determine if interactions within a particular set $V$ of variables contributes to $I_P$, we may check if enforcing the corresponding constraint $C_V$ recovers any additional dependencies not already contained in a *reference MaxEnt distribution*, $P^{(\Omega)}$, constrained by some set of other marginal constraints in $\Omega$. That is, we define the *interaction information*

$$\Delta_\Omega^{(V)} \equiv \left\langle \log_2 \frac{P^{(\Omega \cup V)}}{P^{(\Omega)}} \right\rangle = I_{P^{(\Omega \cup V)}} - I_{P^{(\Omega)}} \equiv I^{(V)} - I^{*(V)}. \tag{1.9}$$

Here, similar to Eq. (1.6), $P^{(\Omega)}$ is the *MaxEnt distribution* satisfying all constraints in $\Omega$ [44], as in

$$P^{(\Omega)} = \frac{1}{Z} \prod_{M \in \Omega} \psi(M) \equiv \frac{1}{Z} \exp\left[-\sum_{M \in \Omega} \varphi(M)\right]. \tag{1.10}$$

By positivity of the Kullback–Leibler divergence, $\Delta^{(V)} \geq 0$. Thus if $\Delta_\Omega^{(V)} > 0$, accounting for the constraint $C_V$ recovers more information, and we say that the *variables in $V$ interact with respect to $P^{(\Omega)}$*.

Note that $\Delta_\Omega^{(V)}$ is $\Omega$-dependent, and to test for dependencies we must first select the reference set of constrained variables $\Omega$. To define an *irreducible interaction* among variables in $V$, we choose $\Omega$ that minimizes the interaction information,

$$\Omega_V = \arg\min_\Omega \Delta_\Omega^{(V)}, \tag{1.11}$$

$$\Delta^{(V)} \equiv \Delta_{\Omega_V}^{(V)}. \tag{1.12}$$

This guarantees that interactions are defined *only* if they cannot be explained away by confounding effects of other statistical dependencies in the network. Then, in particular,

$$I_P \geq \sum_{V \in P(\{X_1,...,X_N\})} \Delta^{(V)}, \quad (1.13)$$

where $P(\{X_1,...,X_N\})$ is the power set (the set of all subset) of the analyzed variables.

**Conjecture 1.** Let $\Omega_1 \subset \Omega_2$ be sets of noncontradictory marginal constraints, and $P^{(\Omega_1)}$ and $P^{(\Omega_2)}$ be the corresponding MaxEnt distributions. Let $V$ be an additional marginal constraint, possibly a subset of either $\Omega_1$ or $\Omega_2$. Then

$$\Delta_{\Omega_1}^{(V)} \geq \Delta_{\Omega_2}^{(V)}. \quad (1.14)$$

Intuitively, this says that interaction informations depend on the order in which the interactions are considered. Dependency bits will be accounted for by the first marginal able to explain them, attributing less bits to later constraints. We have extensively tested this conjecture numerically (not shown), but the proof is not yet available.

According to the Conjecture, the reference set of constraints $\Omega_V$ to test for the existence of *irreducible* interactions within $V$ is

$$\Omega_V = \bigcup_{M \subset P(\{X_1,...,X_N\}), M \not\supseteq V} M. \quad (1.15)$$

Thus $P^{(\Omega_V)}$ preserves all marginals of the original JPD except those that involve all covariates in $V$ simultaneously. This is similar to the Type III Sum of Squares ANOVA for testing significance of predictors. In fact, since $D_{KL}$ is equal to $\chi^2$ asymptotically, the similarity is not accidental. Dependence defined by this choice of $P^{(\Omega_V)}$ is a generalization of conditional dependence with the rest of the network as a condition. This extends the analysis of [10] and defines an interaction among a particular set of variables, rather than within all variable subsets of the same cardinality.

While this formulation gives a precise definition of multivariate statistical dependence, computational issues arise in applying it to large networks. First, searching through the space of all possible multivariate dependencies is exponential in the number of variables as, for $N$ variables, there are $2^N$ possible subsets of the variables. Moreover, each test for an irreducible interaction

$$\Delta^{(V)} > 0 \tag{1.16}$$

requires computing two large MaxEnt distributions, which is not trivial, especially since empirical distributions computed for large-dimensional marginals will be severely undersampled. Finally, in many cases, some of the variables in the network will be unmeasurable (hidden), influencing the interaction structure derivable from the visible variables [10, 46]; this is clearly prevalent in systems biology applications, where we are still far from measuring concentrations of all chemical species in a cell. We will address these issues partially in Sec. 3.

Complications aside, the MaxEnt formulation resolves the problem of disambiguating dynamics arising from different dependency structures, such as in Eqs. (1.1), (1.2). Indeed, independent regulation, as in Eq. (1.1), produces a JPD with only pairwise potentials, while joint regulation requires a third-order potential and will, therefore, result in a third-order interaction.

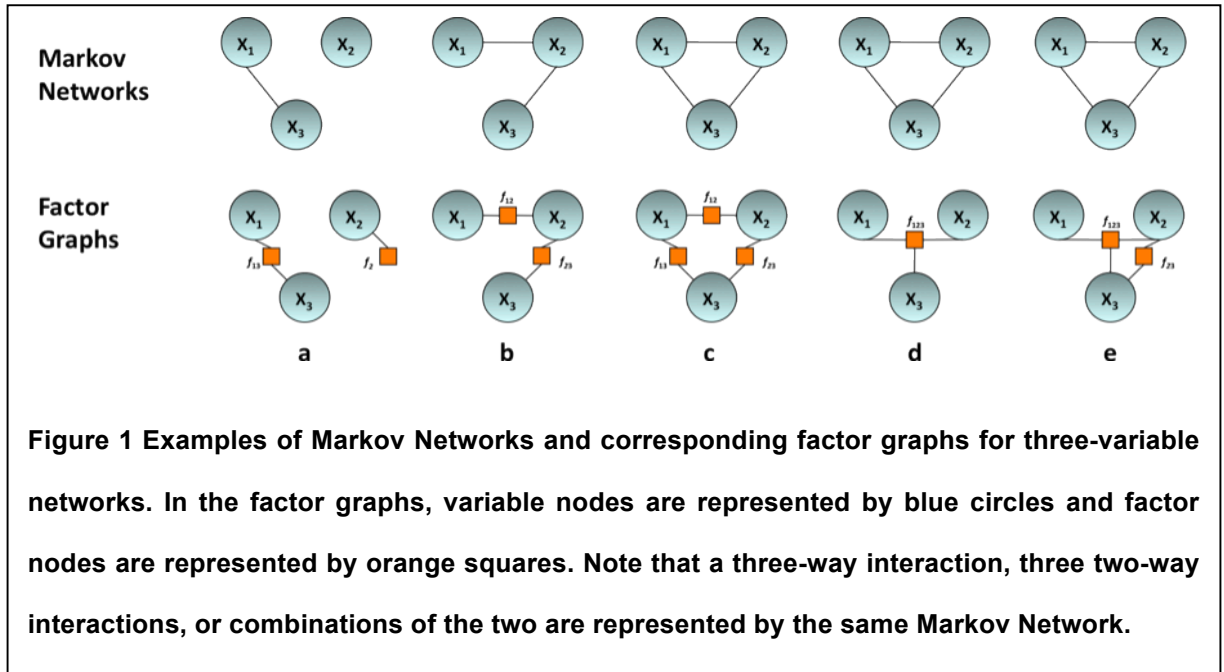

**Figure 1** Examples of Markov Networks and corresponding factor graphs for three-variable networks. In the factor graphs, variable nodes are represented by blue circles and factor nodes are represented by orange squares. Note that a three-way interaction, three two-way interactions, or combinations of the two are represented by the same Markov Network.

## 3. Graphical Models

Graphical models [35] are widely used to provide a visual representation of the factorization of a JPD and to motivate efficient inference algorithms based on graph theoretic considerations. This framework has been applied often in genetic network inference applications [6]. The maximum entropy formulation is strongly related to undirected graphical models. In particular, Eq. (1.10) has the form of a Markov Network, which is visually represented by associating each variable with a node and drawing an edge between each pair of variables that appear together in a potential. However, this network representation is insufficient to distinguish between potentials that are fully parameterized, or only parameterized by functions on subsets of variables, which is a major goal of this work. A more general graphical model, known as a factor graph, is able to represent this distinction. The factor graph representation of a JPD contains two types of nodes. Each factor (potential) $\psi(V)$ is explicitly represented as a *factor node*, with an edge connected to each variable in $V$, which are represented as *variable nodes* (Figure 1). However, in traditional factor graph literature, the factors cannot be defined uniquely once the JPD is known. For example, if a three-variable factor $\psi(X_1,X_2,X_3)$ is present, then any two-variable factor $\psi(X_i,X_j)$, i,j=1,2,3, can be subsumed into it. Put another way, one can set $\psi(X_i,X_j) \to [\psi(X_i,X_j)/f(X_i,X_j)]$ with an arbitrary function $f$, and redefine $\psi(X_1,X_2,X_3) \to [\psi(X_1,X_2,X_3)*f]$ without changing the JPD or its factor graph structure. In particular, setting $f = \psi(X_i,X_j)$ will remove the second order factor completely. Therefore, traditional factor graphs blur the distinction between columns (d) and (e) in Figure 1. Conversely, for the MaxEnt construction of factor graphs, each factor is defined uniquely, so that the factor structure of JPDs in columns (d) and (e) is materially different. Therefore, one can talk about existence or nonexistence of a lower-order factor uniquely and independently of whether the higher-order one involving the same variables exists.

## 3.1 Examples and Properties

We consider a few examples of different graphical model representations for networks of size $M=3$ (larger $M$ is analyzed similarly). First, for a regulatory cascade, or a Markov chain, $X_1 \to X_2 \to X_3$, $P(x_1,x_2,x_3) = P(x_1)P(x_2|x_1)P(x_3|x_2)$, as shown in Figure 1b. Consider the test for $X_1 X_2$ dependence. Following the notation of Eq. (1.9), we let $I^{(12)}$ and $I^{*(12)}$ be the multi-informations of the distributions used to test for dependency on $X_1 X_2$. That is, $\Delta^{(X_1 X_2)} \equiv \Delta^{(12)} = I^{(12)} - I^{*(12)}$. Then, we have $I^{*(12)} = I[X_1,X_3] + I[X_2,X_3] \leq I^{(12)} = I[X_1,X_2] + I[X_2,X_3]$, where the inequality is due to the information processing inequality, and the bound is reached only in special cases. Thus $X_1$, $X_2$ are (generically) dependent. Similarly, $X_2$, $X_3$ are dependent. However, $\Delta^{(13)} = 0$, and $X_1$, $X_3$ are not (even though their marginal mutual information, induced by other interactions, is not zero). Checking for the triplet interactions, we find $I^{*(123)} = I[X_1, X_2] + I[X_2, X_3] = I^{(123)}$, thus no such dependencies are present. If instead $X_2$ regulates $X_1$ and $X_3$, one sees that the dependence structure is the same. Both networks correspond to the graph in Figure 1b.

A more interesting case is when $X_1$, $X_3$ regulate $X_2$ jointly. Here many possibilities exist, not all of them realizable in terms of BN or MRF modeling. First, consider independent regulation: to predict $X_2$, one does not need to know the values of $X_1$ and $X_3$ simultaneously, $P(x_2 | x_1, x_3) = f_{12} f_{23}$, e. g., $P(x_2 | x_1, x_3) \propto \exp\left[-a(x_2 - x_1)^2 - b(x_2 - x_3)^2\right]$ (this corresponds to probabilistic analogues of OR and AND gates [10], to the Lac–repressor [13], and to all regulatory models based on independent binding of transcription factors to the DNA [8]). If $P(X_1 X_3) = P(X_1) P(X_3)$, then the dependency structure is again as in Figure 1b. If in addition there is a regulation $X_1 \to X_3$, so that $P(X_1 X_3) \neq P(X_1) P(X_3)$, then $\Delta^{(13)} \geq 0$, and $\Delta^{(123)} = 0$. The dependency graph now has a loop in it, as in Figure 1c. However, in the case of joint (e.g., cooperative) regulation, $P(X_1 X_2 X_3)$ is nonfactorizable, $\Delta^{(123)} > 0$, and the dependence structure is as in Figure 1d or Figure 1e.

## 3.2 Local Tests

While the previous section described precise tests for three variable networks, computing irreducible statistical dependencies for large networks is computationally intractable. The graphical models framework provides an intuitive interpretation of statistical tests performed on subsets of variables. For example, consider a network with $N \gg 3$ nodes and define $V_3 = \{(1, 2, 3)\}$, and $V_2 = \{(1, 2), (2, 3), (1, 3)\}$. Evaluation of $\Delta^{(V_2)}$ or $\Delta^{(V_3)}$ using Eq. (1.16) is unrealistic since it requires computing MaxEnt distributions with factors over $N-2$ and $N-3$ variables. Instead, one may need to marginalize over many $X_i$, $i > 3$, and search for dependencies in the JPD with 3 variables only. In general, with marginalized (hidden) variables, an irreducible dependency cannot be inferred by MaxEnt tests, but it is informative to understand the meaning of a difference in MaxEnt entropies even in this case.

Due to the factor structure of the JPD in Eq. (1.10), marginalizing over a variable will couple all of its neighbors (nodes with which it participates in a potential) into a single factor. If any of those nodes are marginalized out, its neighbors will further be coupled into this factor, and so on. As a consequence, for any three variables remaining in a marginalized graph, if, in the full factor graph, there exists a factor node such that there is a direct path between it and each of the remaining three variables that does not pass through the other two, then marginalization over hidden variables will produce an effective third-order interaction among the remaining three variables. As discussed in Section 5, this observation has important consequences in genetic network inference and indicates that the proposed multivariate dependency framework can be used to identify proteins that cooperatively interact in a pathway to regulate the expression of a target gene.

## *4. Synthetic data*

A major advantage of our definition of statistical dependencies in terms of the MaxEnt approximations is that it can be applied even when the underlying distributions are undersampled and the corresponding factorizations cannot be readily observed. For $K = \prod K_i$, the cardinality of the JPD[2], larger than the number of samples, $s$, we cannot estimate the distributions reliably, but entropic quantities, and, therefore, the interactions are inferable[3]. Some progress is possible even for $s \sim \sqrt{K}$ [48, 49]. To show this, we used Dirichlet priors [49] to generate random probability distributions with different interaction structures for $N = 3$, and with marginal cardinalities $K_i = 50$. We generated random samples of different sizes, $s = 50...125,000$, from these distributions and tested the quality of inference of the dependencies as a function of $s$. To measure it, we used the *evidence* for an interaction, $E^{(V)} \equiv \Delta^{(V)} / \delta\Delta^{(V)}$, where $\delta\Delta^{(V)}$ is the statistical error of the interaction information. If $E^{(V)}$ is large, the dependency is present. According to Figure 2, proper recovery is possible for $s = K = K_1 K_2 K_3$ *with few assumptions* about the distributions.

---

[2]In genomics, continuous expression levels are routinely discretized. Thus we focus on the discrete case in view of its relevance and conceptual simplicity. Measuring dependencies for continuous variables follows a similar route 47.   Beirlant, J., et al., *Nonparametric entropy estimation: An overview.* Int. J. Math. Stat. Sci., 1997. **6**(1): p. 17--39..

[3]The reader is referred to 47.     Ibid.  and  to  menem.com/ilya/pages/NIPS03  for overviews.

With modern entropy estimation techniques [49], our approach will work even for severely undersampled JPDs. The bottleneck is the estimation of the maximum entropy consistent with the marginals, which currently requires substantial sampling of the marginals, requiring $s \sim \max(K_1 K_2, K_2 K_3, K_1 K_3)$, similarly to the jackknifing method used in [50, 51]. This is encouraging, since the marginals may be well sampled when the JPD is not. However, it is still essential to develop techniques to infer maximum entropies directly. Further, the interaction information is the difference of entropies. It may be small when its error, which is a quadratic sum of the entropy errors, is large. This leads to uncertainties about dependencies even for reliably estimated entropies. Therefore, a method that directly estimates $\Delta$ will be preferred over another entropy–based technique.

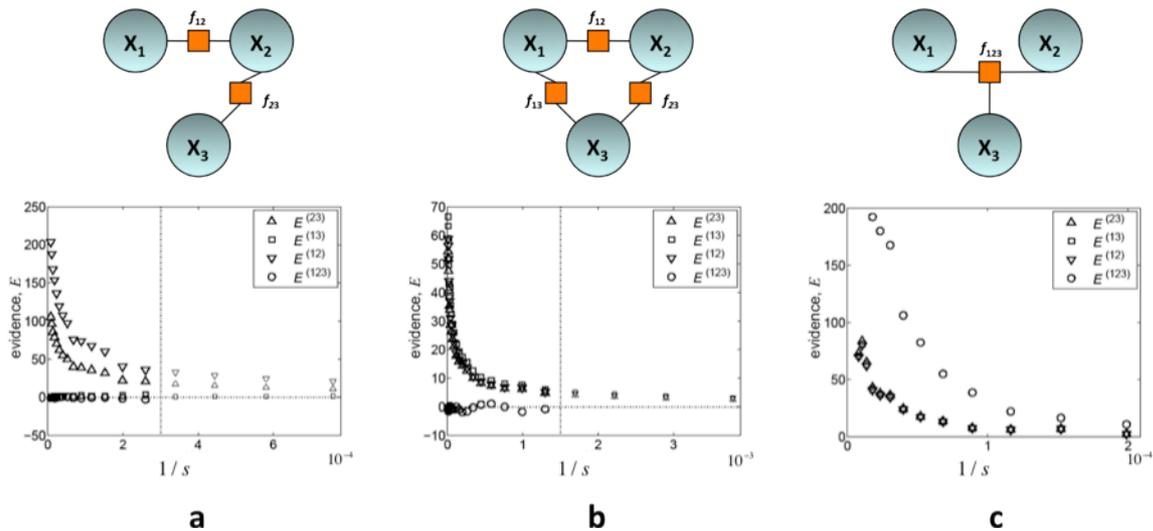

**Figure 2 Inferring regulatory networks from sample size, $s$. We used the NSB [49] method to estimate the entropies (with error bars) of the JPD and its marginals directly. The method does not work for the entropy of $P^{(\Omega)}$ for $\Omega \equiv (123) = \{(X_1, X_2), (X_2, X_3), (X_1, X_3)\}$. Thus IPFP was applied to the counts and the entropy $H_{p^\Omega}$ in the solution was evaluated and extrapolated for $1/s \to 0$ following [50, 51] to account for the sample size dependent bias. The statistical error for each sample size, $s$, was determined by bootstrapping, and the resulting extrapolation**

error was used for $\delta H_{P^{(\Omega)}}$. This approach works since the MaxEnt constraints, like those in Eq. (1.6), are linear in the unknown JPD, $P$, making the biases of $H_P$ and $H_{P^{(\Omega)}}$ behave similarly. Finally, $\Delta^{(V)}$ was calculated as the differences of the appropriate entropies, and $\delta^2 \Delta^{(V)}$ as the sums of squares of the entropy errors. Network models are displayed above each plot. (a) Network with $P \propto \psi(X_1, X_2)\psi(X_2, X_3)$. To the left of the vertical dotted line, $s \approx 3000 > 2^{H_{P^{(\Omega)}}} \ll K \approx 125000$, the sample size corrections are reliable, and all entropies are known well. There is evidence only for $X_1 X_2$ and $X_2 X_3$ interactions, just as it should be. For smaller $s$, the method of [50] fails, but NSB works until $s \sim 2^{1/2 H_P} \approx 60$. For pairwise interactions, we may replace $H_{P^{(\Omega)}}$ by $H_P$ (denoted by smaller markers on the plot) and, since $E^{(13)}$ stays zero nonetheless, and $I(X_1; X_2) + I(X_2; X_3) = I_P$, we still recover the interactions correctly. (b) Network with three pairwise interactions. Again, to the left of the line, $s > 2^{H_{P^{(\Omega)}}}$, all entropies are determined reliably, and there is evidence for all three pairwise interaction, but not for the triplet interaction. To the right of the line, NSB still works, but now we cannot disentangle the loop from the three–way dependence without estimating $H_{P^{(\Omega)}}$. (c) Network with three pairwise and a third-order interaction. Only the regime $s > 2^{H_{P^{(\Omega)}}}$ is shown. The evidence for all three pairwise interactions and for the triplet interaction is barely significant for small $s$ but grows fast.

# 5. Genetic networks inference

## 5.1 Inferring regulatory pathways

The proposed method for identifying multivariate dependencies has important applications for cellular networks inference. Cellular networks are composed of a complex system of interacting and diverse molecular species. For example, consider the task of inferring genetic regulatory interactions using statistical correlations between gene expression array measurements, which measure mRNA concentrations. Generically, genes encode mRNAs, which are translated into proteins. Some of the latter encode transcription factors, which in turn can bind to DNA and influence the expression of other genes. However, mRNA abundance data only probes a small percentage of the regulatory network. For example, the translation of mRNA into protein is dynamically regulated at many levels, including by regulating mRNA stability, nuclear export and cytoplasmic localization, and translation initiation. Once translated, proteins engage in a vast network of interactions, being regulated, for example, by complex formation as well as a variety of post-translational modifications, such as (de)phosphorylation, (de)acetylation, and (de)ubiquitination. Finally, the ability of a gene to be transcribed is strongly affected by modifications of the DNA, such as methylation, chromatin accessibility (which is influenced by histone modifications such as acetylation), as well as other genetic factors including mutations, single nucleotide polymorphisms and chromosomal alterations. Many of these regulatory processes are carried out by proteins, but there is also a critical and ever increasingly appreciated role for other regulatory factors such as non-coding RNAs, metabolites, and extra-cellular signals. The combined effect of these considerations is to create a vast network of hidden variables, while we only probe a small percentage of the system with current technologies. For such complicated regulatory systems, it is difficult to understand the effect of the unobserved variables and thus to interpret the meaning of statistical dependencies between mRNAs.

Section 3.2 provides some insight into this question and suggests that irreducible multivariate statistical dependencies between mRNAs may be used to identify genes that interact in a pathway to jointly regulate the expression of a downstream target. Consider, for example, a transcription factor, TF, that regulates a target gene, T. This interaction is influenced by a (possibly large) number of other proteins, which we call *modulators* [14], denoted by M (Figure 3). The modulators may interact directly with TF, for example via post-transcriptional modification, creating a third-order dependency between TF, M, and T. However, as discussed, effective third-order dependencies are also created between variables that interact indirectly, for example if the modulator regulates another gene or protein that subsequently interacts with TF downstream. This type of series of interactions in which multiple genes jointly control a cellular process (e.g. expression of a target gene) is called a pathway, and is the principle mechanism by which a cell regulates gene expression.

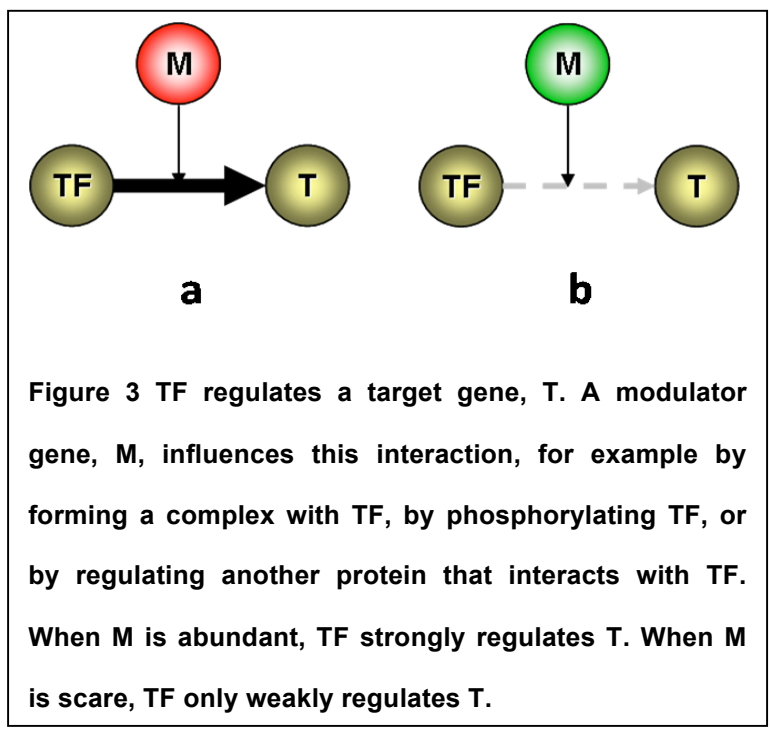

**Figure 3 TF regulates a target gene, T. A modulator gene, M, influences this interaction, for example by forming a complex with TF, by phosphorylating TF, or by regulating another protein that interacts with TF. When M is abundant, TF strongly regulates T. When M is scare, TF only weakly regulates T.**

To identify such third-order dependencies we test for a reduction in entropy by constraining $P(TF,M,T)$ as opposed to constraining $P(TF,T)$, $P(M,T)$, and $P(TF,M)$. The MaxEnt distribution constrained by all three pairwise marginals must be computed by an iterative algorithm. However, a much more computationally efficient procedure can be derived under the simplifying assumption that $TF$ and $M$ are not (irreducibly) statistically dependent, which is a common occurrence in biology [52]. That is, the factorization of the JPD produced by the MaxEnt formalism does not contain the $\psi(TF,M)$ potential. Note that this is less stringent than requiring $I(TF;M)=0$, and only means that we do not need to constrain $(TF, M)$ in the MaxEnt construction. Then the test for the difference in entropy of MaxEnt distributions constrained by $[(TF,T), (M,T)]$ versus that constrained by $[(TF,T,M)]$ reveals a simplified equation based on conditional mutual information. In particular, the MaxEnt distribution constrained by the two pairwise marginals has the form $P^*(TF,T,M) = \frac{P(TF,T)P(M,T)}{P(T)}$ whereas the distribution constrained by the three-way interaction has the form $P^*(TF,T,M) = P(TF,T,M)$. Therefore, letting $\Delta$ denote the difference in multi-information of the two distributions, we have

$$\Delta = H(TF,T) + H(M,T) - H(T) - H(TF,M,T)$$
$$= H(T|M) - H(T|TF,M) + H(TF,T)$$
$$+ H(M) - H(T) - H(M|TF) - H(TF)$$
$$= I(TF;T|M) - I(TF;T) + I(TF;M). \tag{1.17}$$

We implemented this form of the equation, considering cases where $I(TF,M)=0$, ensuring that the simplifying assumption of no statistical interaction between TF and M holds true. This form of the equation was used in [14, 52], but its theoretical basis has not been developed until the current work. This procedure relies on computing whether the mutual information between $TF$ and $T$ increases when conditioned on $M$ under the $I(TF, M) = 0$ assumption. Since expression data are continuous, to overcome the undersampling issue, we use Gaussian kernel estimators for estimating conditional informations [12]. Further, following [52], we discretize $M$ into $M^+$ and $M^-$, representing high and low modulator expressions, and test for $I(TF; T | M^+) \neq I(TF; T | M^-)$ as a proxy for Eq. (1.17). Below we consider how this simplified version of the general framework can be used to identify cooperative regulation, and compare it to using pairwise dependencies only. The main contribution of this work is to formalize the concept of multivariate dependency, and thus we do not claim to exhaustively test its application to biological networks, but rather provide initial evidence of the method's effectiveness for this purpose.

## 5.2  Results for biological networks

The MYC proto-oncogene is a critical regulator of oncogenic onset and progression, and is estimated to be overexpressed in at least one seventh of all human cancers [53], including a large percentage of B cell lymphomas. The pluripotent nature of MYC's interactions make it difficult to characterize the critical pathways that are affected by aberrant MYC expression, and it is thus important to characterize the network of interactions associated with MYC. In addition, MYC provides a convenient test case for reverse engineering algorithms due to a public database containing a large number of biochemically validated MYC transcriptional targets [54]. Moreover, MYC is known to be regulated by the B cell receptor (BCR) pathway in B cells [55], and has over sixty known protein-protein interaction (PPI) partners in the Human Protein Reference Database [56]. Thus, while far from a perfect test, comparing predicted modulators against these two data sources provides a level of validation.

We have recently taken steps towards characterizing the genetic network associated with MYC by analyzing a dataset of 254 microarrays derived from normal and tumor-related human B lymphocyte populations [57]. In particular, we have developed a method [12, 58, 59] that has been used to accurately identify downstream MYC targets [11], and has led to insights into the relationship between the part of the cellular interaction network regulated by MYC, and those regulated by other proto-oncogenes such as NOTCH1 [60]. Further, we have identified a variety of modulators of MYC [14, 52]. In this work we take another, more principled look at the identification of the cellular network that works cooperatively with MYC to jointly regulate sets of target genes.

After filtering out all genes from the microarray exhibiting low expression or insufficient dynamic range, following [52], we compiled two sets of potential modulator genes. The first, which we call signaling molecules (SMs), contains genes that are annotated as protein kinase, protein phosphotase, acetyltransferase or aceylase in the Gene Ontology database, and may potentially post-translationally regulate MYC or another gene that acts in the same pathway as MYC. The second group contains genes with the Gene Ontology annotation of transcription factor (TF) activity, which may serve as co-transcription factors associated with MYC. We also compiled a set of experimentally validated MYC targets from the www.myccancergene.org database [54]. In order to apply Eq. (1.17), we removed potential modulators that had significant MI with MYC, leaving a total of 1,128 Affymetrix probe sets as potential modulators (542 SMs and 598 TFs), which were tested for their ability to modulate MYC interactions with the 340 probe sets associated with MYC targets.

We applied Eq. (1.17) to all combinations of modulators and target genes, with MYC fixed as the TF variable. Statistical significance was assessed using the permutation test described in [52]. This creates a matrix of interaction p-values with all modulators on the columns and all genes on the rows. Significant interactions were defined as those having a Bonferroni corrected p-value less than .05.

We sought to test two specific claims made in the preceding papers [14, 52]. First, that third-order statistics can be used to identify genes that interact in a pathway to indirectly or directly cooperate with a transcription factor to control a set of target genes. Second, that such interactions may be identified by third-order statistics, but not by second-order ones. To this end, we considered all significant third-order interactions and analyzed the number of associated modulators either annotated as belonging to the BCR pathway, or as a protein-protein interaction (PPI) partners with MYC. We call genes meeting these criteria putative modulators. Overall, there were 3,586 and 4,343 significant interactions for the SM and TF datasets, respectively. As shown in Table 1, modulators associated with inferred three-way interactions were significantly enriched with putative modulators[4].

---

[4]We note that p-values are may be overestimated because samples are not independent

|  |  | Putative | Total | Pct | p-value |
|---|---|---|---|---|---|
| **SMs** | All Genes | 12,580 | 74,800 | 16.8% | |
|  | Inferred (three-way) | 1,015 | 3,586 | 28.3% | $3.3*10^{-11}$ |
|  | Inferred (pairwise) | 432 | 3,586 | 12.1% | |
| **TFs** | All Genes | 9,520 | 87,040 | 10.9% | |
|  | Inferred (three-way) | 771 | 4,343 | 17.8% | $4.0*10^{-11}$ |
|  | Inferred (pairwise) | 380 | 4,343 | 8.8% | |

**Table 1 Putative modulators were defined as those contained in the BCR pathway or participating in a PPI with MYC. We considered the percent of putative modulators associated with significant three-way interactions against a background of all tested triplets, as well as triplets with the highest total MI, $I(TF;T)+I(M;T)$. We considered separate statistics for SMs and TFs. Because pairwise statistics identified a lower percent of putative modulators than background, we assessed statistical significance of the third-order interactions against the background. As shown, third-order, but not pairwise, statistics effectively identified putative modulators.**

To test against the hypothesis that modulators can be identified by second-order statistics alone, for each dataset we ranked each interaction based on the total pairwise mutual information, $I(TF;T)+I(M;T)$, and, to compare with third-order tests, considered the top-ranking 3,586 and 4,343 triplets for SMs and TFs, respectively. Only 432 (12.1%) SMs and 380 (8.8%) TFs were putative modulators, indicating that modulators could not be identified using pairwise statistics alone. In fact, the top-ranked interactions based on MI contained a slightly lower percent of putative modulators than the background, likely because the activity of a modulator affects the strength of coupling between the TF and target, diluting the MI. Thus gene triplets with high MI are likely to preferentially not include third-order interactions.

Next, reasoning that important modulators may affect MYC's interaction with a large number of target genes, we tested whether putative modulators could be identified by ranking them based on the number of MYC interactions that they affect. Using this procedure, we can simultaneously identify the modulators of MYC and the lists of target genes that they modulate. ROC analysis (Figure 4a) showed that the top-ranking genes by this procedure were significantly enriched for putative modulators. The top-ranking gene, casein kinase 2 alpha 1 (CSNK2A), showed a clear and strong pattern of positive modulation of MYC (Figure 4b,c), and has been experimentally validated in vivo to directly phosphorylate MYC and positively modulate its DNA binding kinetics [61, 62]. Finally, the binding sites for the top-ranking TF modulators were significantly enriched in the promoters of inferred target genes (Figure 4d), providing evidence that these co-transcription factors cooperate with MYC by binding to the promoters of common targets. Together, these results indicated that this procedure could effectively identify genes that interact in a cellular pathway of interest.

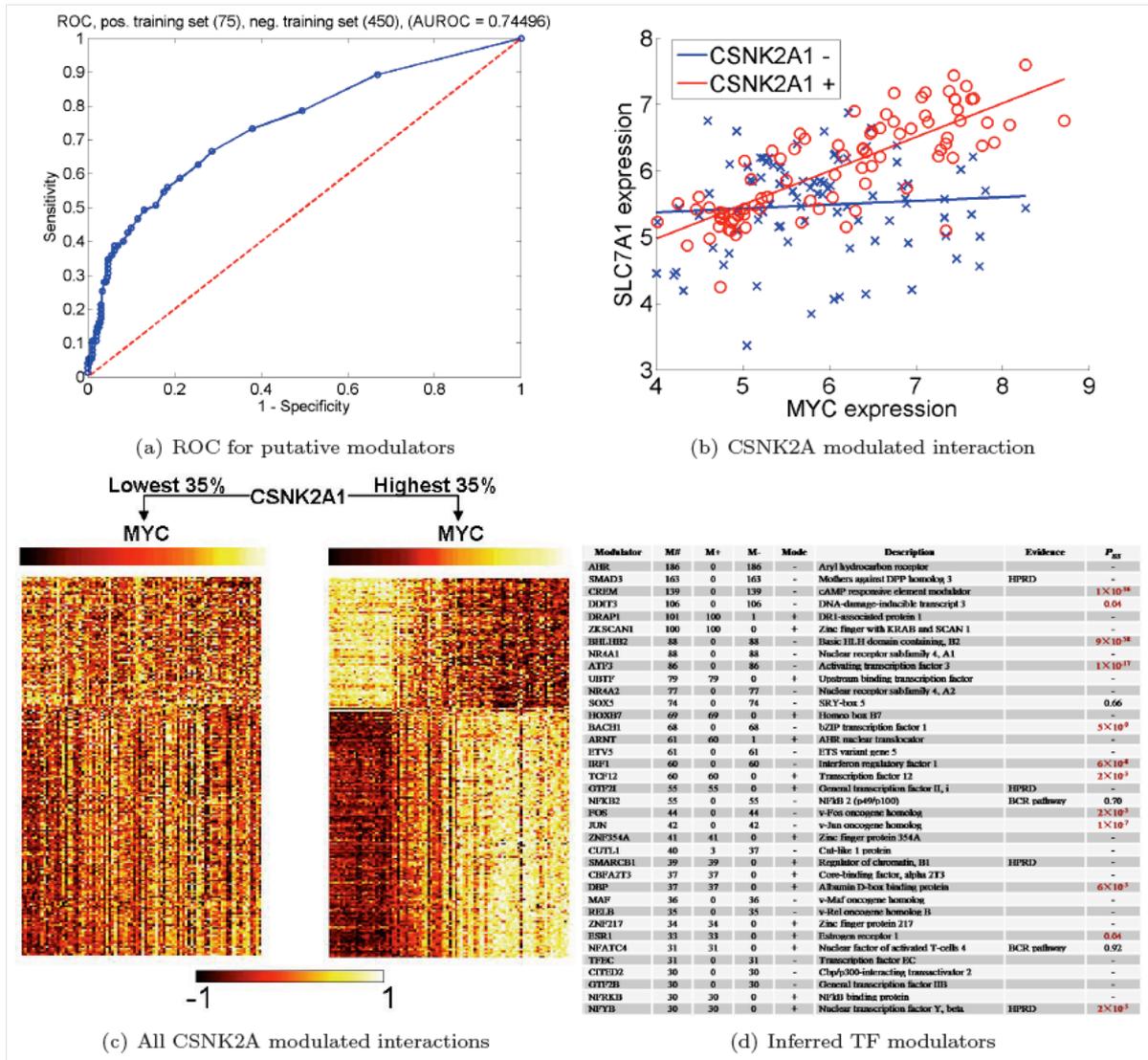

**Figure 4** (a) A set of 75 putative modulators was compiled, including probe sets from BCR pathway genes and known PPI partners of MYC, together with 450 negative instances, including those not in the positive set and not correlated with any probe sets in the positive set. Each probe set was ranked based on the number of MYC interactions that it modulated. By varying this number as the threshold, a Receiver Operating Characteristic (ROC) curve was produced. The area under the curve was calculated to be 0.74. (b) Example scatter plot of an interaction modulated by CSNK2A1, a bona-fide positive modulator of MYC. Expression levels (in log) of MYC and SLC7A1 (a known MYC target) were plotted on the X and Y-axes, respectively. Data was partitioned into the 35% of samples with the highest CSNK2A1 expression and the 35% of samples with the lowest CSNK2A1 expression (circles and crosses, respectively), and a line was fitted to the data points in each subset. As shown, when

CSNK2A1 was highly expressed, MYC strongly regulated SLC7A1, whereas this interaction was not apparent at low CSNK2A1 expression levels. (c) MYC target gene expression modulated by CSNK2A1. Two microarray images (modulated MYC target genes on rows and samples on columns) are shown for each subset of high and low CSNK2A1 expression. Samples in each subset were sorted by MYC expression and expression values of target genes were rank transformed, scaled between -1 and 1, and displayed using the color scheme indicated by the colorbar shown at the bottom of the plot. At high CSNK2A1 expression, MYC was highly correlated with these targets, but not at low CSNK2A1 expression. (d) TF binding site enrichment analysis for MYC modulators functioning as potential co-transcription factors. For each modulator with an available scoring matrix in the TRANSFAC database [63], its binding sites were searched for in the promoter regions (2K upstream and 2K downstream from transcription start site) of each modulated MYC target gene. Binding site enrichment for each modulator was assessed using Fisher's exact test and comparing to 13,000 random human promoters. M#: number of modulated MYC targets; M+/-: number of MYC targets positively/negatively affected by the modulators; $P_{BS}$: p-value of the binding site enrichment test. Twelve of the top fifteen inferred co-transcription factor modulators that had available scoring matrices in TRANSFAC displayed statistically significant enrichment of their DNA binding site in the promoters of the inferred target genes.

# 6. Conclusions

In this article, we have revisited the concept of multivariate dependence using information theoretic, maximum-entropy based techniques. We have provided a definition of a higher-order statistical interaction that is able to measure the interaction strength, in bits, and to assign it to a specific set of statistical co-variates. This extended earlier results of Schneidman et al. [10], which allowed for identification of the existence of a higher-order interaction, but could not identify which specific variables participated in it.

As with every definition, ours is useful only to the extent that it can be applied in practical situations. To verify this, we explored how identification of multivariate dependencies is affected by undersampling that is typical of real-life problems. Further, we argued that the definition allows us to take a new, principled look at reverse-engineering of transcriptional regulatory networks, in particular on identification of combinatorially regulated pathways in transcriptional data. To promote the suitability of the method, we designed a proxy test that well-approximates our definition of multivariate dependence in typical transcriptional regulation data. The test allowed for clear interpretation of synthetic gene expression data, and it made specific, verifiable, and literature-supported predictions about regulatory cofactors, also called modulators, operating together with MYC to regulate its targets.

Clearly, the method is still in the early stages of development. To complete the definition, the Conjecture that allowed us to define the interaction information uniquely needs to be proved. Further, for applications, development of techniques for dealing with undersampling for identification of higher-order dependencies is likely the largest obstacle to a wide adoption of the method. Finally, additional testing is required to validate the applicability of the approximate test to various biological data. We will return to all of these questions in future work. However, in its present form, we believe that the definition of multivariate dependence introduced in this work provides an important theoretical advance in the field of statistical inference, with applications to systems biology and related disciplines.

## 7. Acknowledgements

A.A.M. was supported by an IBM Ph.D. fellowship during earlier stages of this work. I.N. acknowledges support from Columbia University, KITP/UCSB, DOE under Contract No. DE-AC52-06NA25396, and NSF under Grant No. ECS-0425850 during earlier stages of this work.

# 8. References

1. Margolin, A.A., *Computational inference of genetic networks in human cancer cells.* 2009, Saarbrucken, Germany: VDM Verlag. 288.
2. Schena, M., et al., *Quantitative monitoring of gene expression patterns with a complementary DNA microarray.* Science, 1995. **270**(5235): p. 467-70.
3. Lu, J., et al., *MicroRNA expression profiles classify human cancers.* Nature, 2005. **435**(7043): p. 834-8.
4. Perez, O.D. and G.P. Nolan, *Simultaneous measurement of multiple active kinase states using polychromatic flow cytometry.* Nat Biotechnol, 2002. **20**(2): p. 155-62.
5. Lu, W., E. Kimball, and J.D. Rabinowitz, *A high-performance liquid chromatography-tandem mass spectrometry method for quantitation of nitrogen-containing intracellular metabolites.* J Am Soc Mass Spectrom, 2006. **17**(1): p. 37-50.
6. Friedman, N., *Inferring cellular networks using probabilistic graphical models.* Science, 2004. **303**(5659): p. 799-805.
7. Beer, M.A. and S. Tavazoie, *Predicting gene expression from sequence.* Cell, 2004. **117**(2): p. 185-98.
8. Bussemaker, H.J., H. Li, and E.D. Siggia, *Regulatory element detection using correlation with expression.* Nat Genet, 2001. **27**(2): p. 167-71.
9. Slonim, N., O. Elemento, and S. Tavazoie, *Ab initio genotype-phenotype association reveals intrinsic modularity in genetic networks.* Mol Syst Biol, 2006. **2**: p. 2006 0005.
10. Schneidman, E., et al., *Network information and connected correlations.* Phys Rev Lett, 2003. **91**(23): p. 238701.
11. Basso, K., et al., *Reverse engineering of regulatory networks in human B cells.* Nat Genet, 2005. **37**(4): p. 382-90.
12. Margolin, A.A., et al., *ARACNE: an algorithm for the reconstruction of gene regulatory networks in a mammalian cellular context.* BMC Bioinformatics, 2006. **7 Suppl 1**: p. S7.
13. Buchler, N.E., U. Gerland, and T. Hwa, *On schemes of combinatorial transcription logic.* Proc Natl Acad Sci U S A, 2003. **100**(9): p. 5136-41.
14. Wang, K., et al., *Genome-wide identification of post-translational modulators of transcription factor activity in human B cells.* Nat Biotechnol, 2009. **27**(9): p. 829-39.
15. de la Fuente, A., et al., *Discovery of meaningful associations in genomic data using partial correlation coefficients.* Bioinformatics, 2004. **20**(18): p. 3565-74.
16. Hartemink, A.J., et al., *Using graphical models and genomic expression data to statistically validate models of genetic regulatory networks.* Pac Symp Biocomput, 2001: p. 422-33.
17. Pe'er, D., et al., *Inferring subnetworks from perturbed expression profiles.* Bioinformatics, 2001. **17 Suppl 1**: p. S215-24.
18. Anastassiou, D., *Computational analysis of the synergy among multiple interacting genes.* Mol Syst Biol, 2007. **3**: p. 83.
19. Watkinson, J., et al., *Inference of regulatory gene interactions from expression data using three-way mutual information.* Ann N Y Acad Sci, 2009. **1158**: p. 302-13.
20. Agresti, A., *Categorical data analysis*. 1990, New York: Wiley. xv, 558 p.

21. Joe, H., *Multivariate models and dependence concepts*. 1997, Boca Raton: Chapman and Hall.
22. Darroch, J.N., *Interactions in multi-factor contingency tables.* J. Roy. Stat. Soc. Ser. B (Methodol.), 1962. **24**(1): p. 251--263.
23. Lancaster, H.O., *Complex contingency tables treated by the partition of chi square.* J. Roy. Stat. Soc. Ser. B (Methodol.), 1951. **13**(2): p. 242--249.
24. Roy, S.N. and M.A. Kastenbaum, *On the hypotheis of no "interaction" in a multi-way contingency table.* Ann. Math. Stat., 1956. **27**(3): p. 749--757.
25. Cover, T.M. and J.A. Thomas, *Elements of information theory*. Wiley series in telecommunications. 1991, New York: Wiley. xxii, 542 p.
26. Shannon, C.E., *A Mathematical Theory of Communication.* Bell System Technical Journal, 1948. **27**(3): p. 379-423.
27. Ireland, C.T. and S. Kullback, *Contingency tables with given marginals.* Biometrika, 1968. **55**(1): p. 179--188.
28. Kullback, S., *Probability densities with given marginals.* Ann. Math. Stat., 1968. **39**(4): p. 1236--1243.
29. Bell, A.J., *Co-Information Lattice.* Tech. Report RNI-TR-02-1, 2002.
30. Chechick, G. *Groups redundancy measures reveal redundancy reduction along the auditory pathway*. in *Adv. Neural Inf. Proc. Syst. 14*. 2002. Cambridge, MA: MIT Press.
31. Garner, W.R. and W.J. McGill, *The relation between information and variance analysis.* Psychometrika, 1956. **21**(3): p. 219--228.
32. Good, I.J., *Maximum entropy for hypothesis formulation, especially for multidimensional contingency tables.* Ann. Math. Stat., 1973. **34**(3): p. 911--934.
33. McGill, W., *Multivariate information transmission.* IRE Trans. Inf. Thy., 1954. **4**: p. 93-110.
34. Watanabe, S., *Information Theoretical Analysis of Multivariate Correlation.* IBM J. of Research and Development, 1960. **4**(1): p. 66--82.
35. Pearl, J., *Probabilistic reasoning in intelligent systems: networks of plausible inference*. 1988, San Francisco, CA: Morgan Kaufmann Publishers, Inc.
36. Jakulin, A. and I. Bratko, *Quantifying and Visualizing Attribute Interactions.* arXiv:cs/0308002, 2003.
37. Kullback, S. and R.A. Leibler, *On Information and Sufficiency.* Annals of Mathematical Statistics, 1951. **22**(1): p. 142-143.
38. Jaynes, E.T., *Information Theory and Statistical Mechanics.* Phys. Rev., 1957. **106**: p. 620--630.
39. Lewis, P.M., *Aproximating probability distributions to reduce storage requirements.* Information and Control, 1959. **2**: p. 214--225.
40. Ku, H.H. and S. Kullback, *Interaction in multidimensional contingency tables: an information theoretic approach.* J. Res. Natl. Bur. Stand. (Math. Sci), 1968. **72B**(3): p. 159--200.
41. Amari, S., *Information Geometry on Hierarchy of Probability Distributions.* IEEE Trans. Inf. Thy., 2001. **47**(5): p. 1701--1711.
42. Martignon, L., *Neural Coding: Higher-Order Temporal Patterns in the Neurostatistics of Cell Assemblies.* Neural Comput., 2000. **12**: p. 2621--2653.
43. Soofi, E., *A generalized formulation of conditional logit with diagnostics.* J. Amer. Stat. Assoc., 1992. **87**(419): p. 812--816.
44. Csiszar, I., *I-divergence geometry of probability distributions and minimization problems.* Ann. Probab., 1975. **3**(1): p. 146--158.

45. Deming, W.E. and F.F. Stephan, *On a least squares adjustment of a sampled frequency table when the expected marginal totals are known.* Annals of Mathematical Statistics, 1940. **11**: p. 427-444.
46. Margolin, A.A. and A. Califano, *Theory and limitations of genetic network inference from microarray data.* Ann N Y Acad Sci, 2007. **1115**: p. 51-72.
47. Beirlant, J., et al., *Nonparametric entropy estimation: An overview.* Int. J. Math. Stat. Sci., 1997. **6**(1): p. 17--39.
48. Nemenman, I., W. Bialek, and R.d.R.v. Steveninck, *Entropy and information in neural spike trains: Progress on the sampling problem.* Phys. Rev. E, 2004. **69**(5): p. 056111.
49. Nemenman, I., F. Shafee, and W. Bialek, *Entropy and Inference, Revisited*, in *Advances in Neural Information Processing Systems 14*, T.G. Dietterich, S. Becker, and Z. Ghahramani, Editors. 2002: Cambridge, MA.
50. Strong, S.P., et al., *Entropy and information in neural spike train.* Phys. Rev. Lett., 1998. **80**: p. 197-200.
51. Nemenman, I., et al., *Neural coding of natural stimuli: information at sub-millisecond resolution.* PLoS Comput Biol, 2008. **4**(3): p. e1000025.
52. Wang, K., et al. *Genome-wide identification of modulators of cellular networks in human B lymphocytes*. in *Proceedings of the 10th Annual International Conference on Research in Computational Molecular Biology (RECOMB)*. 2006.
53. Fernandez, P.C., et al., *Genomic targets of the human c-Myc protein.* Genes Dev, 2003. **17**(9): p. 1115-29.
54. Zeller, K.I., et al., *An integrated database of genes responsive to the Myc oncogenic transcription factor: identification of direct genomic targets.* Genome Biol, 2003. **4**(10): p. R69.
55. Niiro, H. and E.A. Clark, *Regulation of B-cell fate by antigen-receptor signals.* Nat Rev Immunol, 2002. **2**(12): p. 945-56.
56. Peri, S., et al., *Development of human protein reference database as an initial platform for approaching systems biology in humans.* Genome Res, 2003. **13**(10): p. 2363-71.
57. Klein, U., et al., *Gene expression profiling of B cell chronic lymphocytic leukemia reveals a homogeneous phenotype related to memory B cells.* J Exp Med, 2001. **194**(11): p. 1625-38.
58. Margolin, A.A., et al. *On the reconstruction of interaction networks with applications to transcriptional regulation*. in *Neural Information Processing Systems*. 2004. Whistler, BC, Canada.
59. Margolin, A.A., et al., *Reverse Engineering Cellular Networks.* Nature Protocols, 2006. **1**(2): p. 662-671.
60. Palomero, T., et al., *NOTCH1 directly regulates MYC and activates a feed-forward-loop transcriptional network promoting leukemic cell growth.* Proc Natl Acad Sci, 2006. **103**(48).
61. Luscher, B., et al., *Myc oncoproteins are phosphorylated by casein kinase II.* Embo J, 1989. **8**(4): p. 1111-9.
62. Bousset, K., et al., *Identification of casein kinase II phosphorylation sites in Max: effects on DNA-binding kinetics of Max homo- and Myc/Max heterodimers.* Oncogene, 1993. **8**(12): p. 3211-20.
63. Matys, V., et al., *TRANSFAC: transcriptional regulation, from patterns to profiles.* Nucleic Acids Res, 2003. **31**(1): p. 374-8.